\documentclass[conference]{IEEEtran}

\usepackage[pdftex]{graphicx}
\graphicspath{{./images/}}

\usepackage[cmex10]{amsmath}
\usepackage{amssymb}

\title{A GPU-based Correlator X-engine\\ Implemented on the CHIME Pathfinder}

\author{\IEEEauthorblockN{
		Nolan Denman,\IEEEauthorrefmark{1}\IEEEauthorrefmark{2} 
		Mandana Amiri,\IEEEauthorrefmark{4}
		Kevin Bandura,\IEEEauthorrefmark{3} 
		Jean-Fran\c{c}ois Cliche,\IEEEauthorrefmark{3}
		Liam Connor,\IEEEauthorrefmark{1}\IEEEauthorrefmark{2}\IEEEauthorrefmark{5}\\
		Matt Dobbs,\IEEEauthorrefmark{3}\IEEEauthorrefmark{6}
		Mateus Fandino,\IEEEauthorrefmark{4}
		Mark Halpern,\IEEEauthorrefmark{4}
		Adam Hincks,\IEEEauthorrefmark{4}
		Gary Hinshaw,\IEEEauthorrefmark{4}
		Carolin H{\"{o}}fer,\IEEEauthorrefmark{4}\\
		Peter Klages,\IEEEauthorrefmark{1} 
		Kiyoshi Masui,\IEEEauthorrefmark{4}\IEEEauthorrefmark{6}
		Juan Mena Parra,\IEEEauthorrefmark{3} 
		Laura Newburgh,\IEEEauthorrefmark{1}
		Andre Recnik,\IEEEauthorrefmark{1}
		J. Richard Shaw,\IEEEauthorrefmark{5}\\
		Kris Sigurdson,\IEEEauthorrefmark{4}
		Kendrick Smith,\IEEEauthorrefmark{7}
		and
		Keith Vanderlinde\IEEEauthorrefmark{1}\IEEEauthorrefmark{2}
	}
\IEEEauthorblockA{\IEEEauthorrefmark{1}
		Dunlap Institute,
		University of Toronto
	}
\IEEEauthorblockA{\IEEEauthorrefmark{2}
		Department of Astronomy and Astrophysics,
		University of Toronto
	}
\IEEEauthorblockA{\IEEEauthorrefmark{4}
		Department of Physics and Astronomy, University of British Columbia
	}
\IEEEauthorblockA{\IEEEauthorrefmark{3}
		Department of Physics, McGill University
	}
\IEEEauthorblockA{\IEEEauthorrefmark{5}
		Canadian Institute for Theoretical Astrophysics
	}
\IEEEauthorblockA{\IEEEauthorrefmark{6}
		Canadian Institute for Advanced Research, CIFAR Program in Cosmology and Gravity
	}
\IEEEauthorblockA{\IEEEauthorrefmark{7}
		Perimeter Institute for Theoretical Physics
	}
	Contact E-Mail: denman@astro.utoronto.ca
}

\begin{document}

\bstctlcite{BSTcontrol}

\maketitle

\begin{abstract}
We present the design and implementation of a custom GPU-based compute cluster 
that 
provides the correlation X-engine of the CHIME Pathfinder radio telescope.
It is among the largest such systems in operation, correlating 32,896 baselines (256 inputs) over 400\,MHz of radio bandwidth.
Making heavy use of consumer-grade parts and a custom software stack, the system was developed at a small fraction of the cost of comparable installations.
Unlike existing GPU backends, this system is built around OpenCL kernels running on consumer-level AMD GPUs, taking advantage of low-cost hardware and leveraging packed integer operations to double algorithmic efficiency.
The system achieves the required 105\,TOPS in a 10\,kW power envelope, making it one of the most power-efficient X-engines in use today.
\end{abstract}

\section{Introduction}
\label{sect:intro}

The Canadian Hydrogen Intensity Mapping Experiment (CHIME) is an interferometric radio telescope, presently under construction at the Dominion Radio Astrophysical Observatory (DRAO) in British Columbia, Canada,
which will map the northern sky over a radio band from 400 to 800\,MHz.
With over 2000 inputs and a 400\,MHz bandwidth, the correlation task (measured as the bandwidth-baselines product) on CHIME will be an order-of-magnitude larger than on any currently existing telescope array.
The correlator follows an FX split design, 
with a first stage Field Programmable Gate Array (FPGA)-based F-engine which digitizes, Fourier transforms (channelizes), and bundles the data into independent frequency bands, 
followed by a second-stage Graphics Processing Unit (GPU)-based X-engine which 
produces a spatial correlation matrix consisting of the integrated pairwise products of all the inputs at each frequency.

The CHIME Pathfinder instrument \cite{chime-pf} 
features 128 dual-polarization feeds 
and a reduced-scale prototype of the full CHIME correlator.
This paper describes the X-engine of the Pathfinder's 256-input hybrid FPGA/GPU FX correlator,
among the largest such systems in operation.
Through extensive use of off-the-shelf consumer-grade hardware and heavy optimization of custom data handling and processing software, the system achieves high performance at a small fraction of the hardware cost of comparable installations.

This paper focuses on the system architecture and implementation, while
two companion papers describe the custom software stacks, one focusing on an innovative OpenCL-based X-engine GPU kernel \cite{klages}, and one on the handling of the vast data volume flowing through the system \cite{recnik}.

The paper is structured as follows:
design considerations and constraints are discussed in \S\ref{sec:des-con};
the hardware components of the system are described in \S\ref{sec:hard}, and the software in \S\ref{sec:soft};
the scaling of the X-engine to the full-size CHIME telescope is described in \S\ref{sec:scale}, and a summary and conclusion follow in \S\ref{sec:conc}.

\section{Design Considerations}
\label{sec:des-con}
While most components in CHIME scale linearly with number of inputs $N$,
the computational cost of pairwise correlation scales as $N^2$,
making efficiency in the X-engine a primary concern.
There are correlation techniques which rely on the redundancy of CHIME feed
separations to scale as $N\log N$, but the real-time calibrations these require 
for precision observations remain largely unproven in an astrophysical context.
Design decisions were guided by the need to produce an inexpensive system 
capable of scaling to support full CHIME,
and which would support rapid development and deployment of new data processing algorithms.
These requirements of computational power and ease of development
drove the decision to build the X-engine around GPUs rather than Application-Specific Integrated Circuits (ASICs) or FPGAs.

The computational cost $\eta$ of pairwise element correlation for $N$ elements across a bandwidth of $\Delta\nu$ is
\begin{equation}\eta = \Delta\nu \cdot N \cdot (N+1) / 2\end{equation}
measured in complex multiply-accumulate (cMAC) operations per second;
for the CHIME Pathfinder, $\eta=13$\,TcMAC/s.
For large $N$ this dominates the cost of any other processing proposed for the X-engine.
Top-end GPUs in 2014 provided of order 4$\,$TFLOPS of processing power per chip, equivalent to 0.5\,TcMAC/s;
a na\"{\i}ve but efficient  X-engine implementation for the CHIME Pathfinder would require of order 26 GPUs, although network topology considerations favor a baseline target of 32 GPUs in the cluster.

Several factors favor a densely-packed configuration for the X-engine.
The proximity of the Pathfinder correlator system to the telescope itself, in a national radio-quiet zone,
requires substantial Faraday shielding around the entire system, the cost of which increases rapidly with size.
The 10\,GbE 4xSFP+\,$\leftrightarrow$\,QSFP+ cabling chosen for the Pathfinder system shows a steep increase in cost above a length of 7\,m, where active cabling becomes necessary.
Together, these apply substantial pressure to minimize the physical dimensions of the correlator system.
The nodes hosting the GPUs also form the bulk of the hardware cost,
so maximizing the density of GPUs within each node is an efficient cost control measure.

\subsection{I/O Requirements}
Data arrive from the F-engine as 4+4-bit offset-encoded\footnote{That is, a range of [-8,7] is set to [0,15] by adding 8 to the values.} complex numbers, arranged into 1024 independent frequency sub-bands.
Assuming the baseline 32-GPU layout, each GPU is independently processing 32 of these bands,
corresponding to a 12.5\,MHz band of radio data.
In this layout each GPU requires a continuous 25.6\,Gbps input stream of sky data drawn from four 10\,GbE connections.

If limited to standard bus widths, 8-lane PCI Express Revision 3 becomes the baseline interconnect,
meaning 16 lanes of PCIe3 are required for each of the 32 GPUs -- 
8 to feed the GPU, and 8 to receive data from a NIC.
While PCIe3x4 or PCIe2x8 links could theoretically support this transfer rate,
tests showed it to be unreliable and prone to lost packets and bottlenecking.
The use of PCIe3 interconnects restricts the choice of CPU to Intel varieties, and
the Ivy Bridge line of consumer processors provide up to 40 lanes of PCIe3,
allowing each host system to support two GPUs, fed from eight 10\,GbE links.

The rate at which correlated output data are produced depends linearly on the choice of accumulation period, but is not a significant driver in the design;
for the Pathfinder system's default 21\,s integration, the output is a modest $\sim$8\,Mb/s per node.

\subsection{GPU Selection}
Consumer-grade hardware was generally preferred to professional for its increased availability, interoperability of components, and drastically reduced cost.
Dedicated scientific GPUs are differentiated by their use of error-correcting memory and the availability of native double-precision functions,
neither of which are significant considerations in a radio correlator.
The AMD offerings feature higher computational throughput per unit cost than the comparable NVIDIA GPUs, which led to an early preference for AMD-brand consumer-grade GPUs.
This is a significant departure from prior GPU X-engines, which typically use the xGPU \cite{xGPU} software package, a highly optimized X-engine implementation built on NVIDIA's proprietary CUDA programming framework.
A custom software X-engine was developed in its place (\S\ref{sec:crosscorr}, \cite{klages}),
built on the vendor-independent OpenCL standard. \cite{opencl}

Proof-of-concept software was developed on 2012-era AMD Southern Islands GPUs similar to those later adopted for the CHIME Pathfinder.
This software was able to process all 32,896 baselines of simulated Pathfinder data at the required 12.5\,MHz/GPU rate.
Further software optimization brought the per-GPU throughput up to $\sim$19\,MHz, leaving ample headroom for additional processing (\S\ref{sec:gpu_algs}).

Consumer-grade GPU boards are manufactured by a variety of third parties,
with proprietary modifications to the reference design which can result in
higher or lower power draw, faster or slower default clocking,
variable power-saving or high-temperature operation; 
and which vary dramatically in their solutions for heat dissipation.
The cooling systems are generally not designed for the close packing required for the CHIME Pathfinder:
of the 6 varieties of R9 280X boards tested in a packed 4U case, many reached die temperatures close to 100$^\circ$C.
Operation at these temperatures lowers the GPUs' efficiency and increases their failure rate while drawing additional power for cooling fans.
Table \ref{tab:salient} shows the maximum die temperature, mean power draw, and processing throughput
of GPUs running an unoptimized `stress test' kernel in a closely-packed case similar to those used in the Pathfinder X-engine.
These measurements were performed at an ambient air temperature of $\sim$25$^\circ$C,
and differences in power draw and throughput are attributable to a combination of changes in clocking, efficiency, and variable fan speeds.

The best performers were Sapphire-branded ``Dual-X OC'' boards,
which maintained average die temperatures of $\sim$75$^\circ$C in the test configuration,
while matching the computational performance of the other boards and consuming less power.
Other strong performers (e.g., the HIS-branded boards) showed higher
pricing markups and proved more difficult to source.

\begin{table}[!t]
\renewcommand{\arraystretch}{1.3}
\centering
\caption{Operational thermal parameters of dense-packed \break AMD R9 280X GPUs from different manufacturers}
\label{tab:salient}
\begin{tabular}[b]{|l|c|c|c|} \hline
	{\bf Manufacturer} & {\bf Max T ($^\circ$C)} & {\bf Power (W)} & {\bf BW (MHz)}\\\hline
	MSI OC Edition	& $>$99 & 825 	& 13.6\\
	XFX 			& $>$99 & 800 	& 13.0\\
	Club3D RoyalQueen & 96 	& 875	& 14.2\\
	TurboDuo 		& 96 	& 800	& 13.6\\
	HIS IceQ 		& 89 	& 730	& 14.5\\
	Sapphire Dual-X OC & 85 & 715	& 13.5\\
	Sapphire, Watercooled & T$_\textrm{water}$ + 10$^\circ$C & 510 & 14.5\\\hline
\end{tabular}
\end{table}

\section{Implementation - Hardware}
\label{sec:hard}
\begin{figure}[!t]
	\centering
	\includegraphics[width=3.2in]{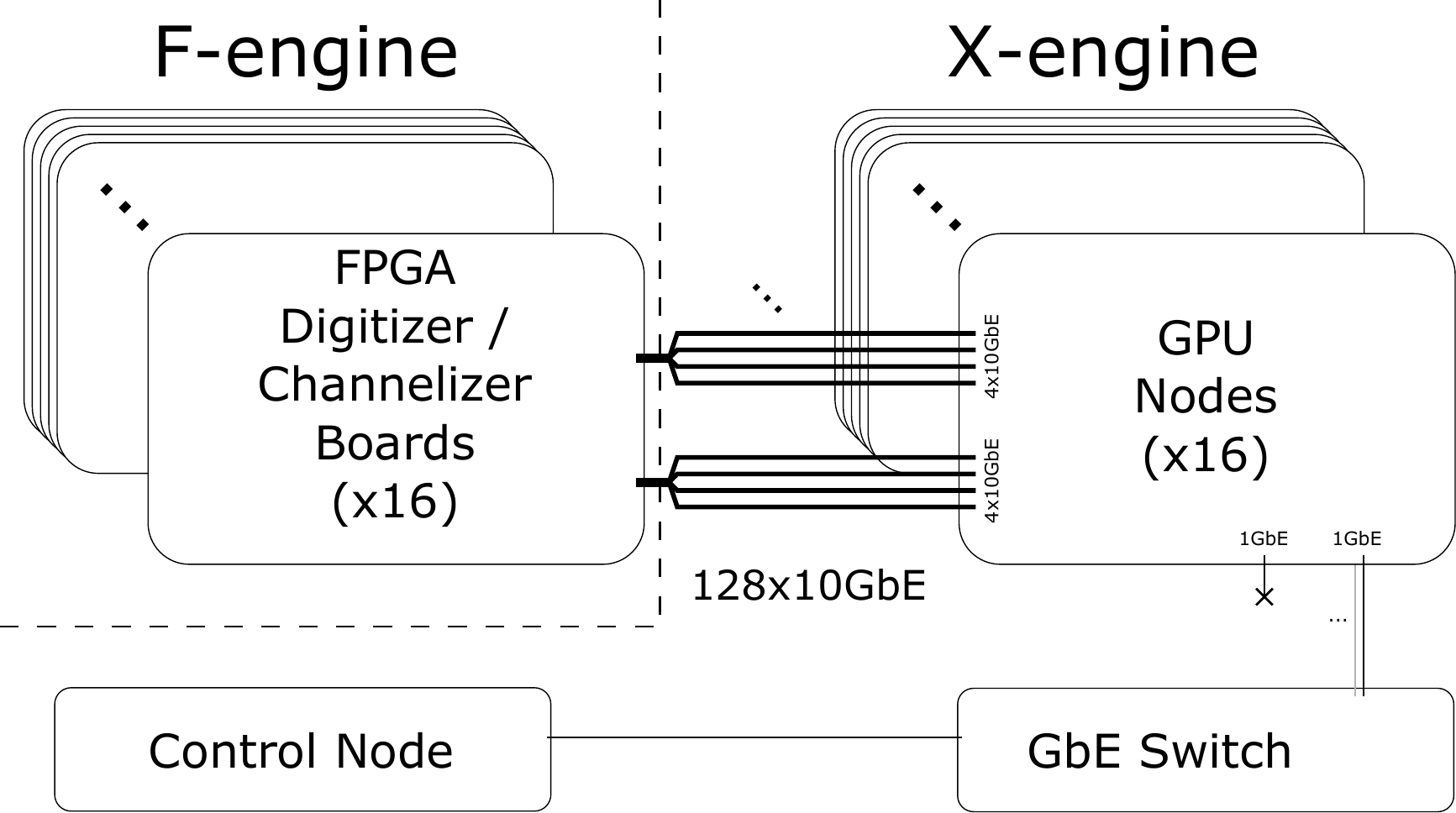}
	\caption{A schematic of the data flow between the F- and X-engine portions of the correlator. The interchange switch and the {\tt gamelan} control node are also shown.}
	\label{fig:cluster}
\end{figure}

The X-engine is a cluster of 16 nodes, each housed in a 4U case.
It fits in 64U spread across two 42U racks, dubbed East and West, containing 6 and 10 nodes respectively.
Each rack is powered by a Tripp~Lite PDU3VSR10H50 managed PDU, allowing remote monitoring and power cycling of all nodes.
The remaining space in the East rack contains the FPGA-based F-engine as well the 1GbE interchange switch,
while the West rack additionally houses a 1U control system dubbed {\tt gamelan}.\footnote{An Indonesian percussion ensemble featuring chimes; see http://en.wikipedia.org/wiki/Gamelan}
A diagram of the X-engine computational cluster is shown in Figure \ref{fig:cluster}.

\subsection{Node Description}
\label{sec:node}
\begin{figure}[!t]
	\centering
	\includegraphics[width=2.5in]{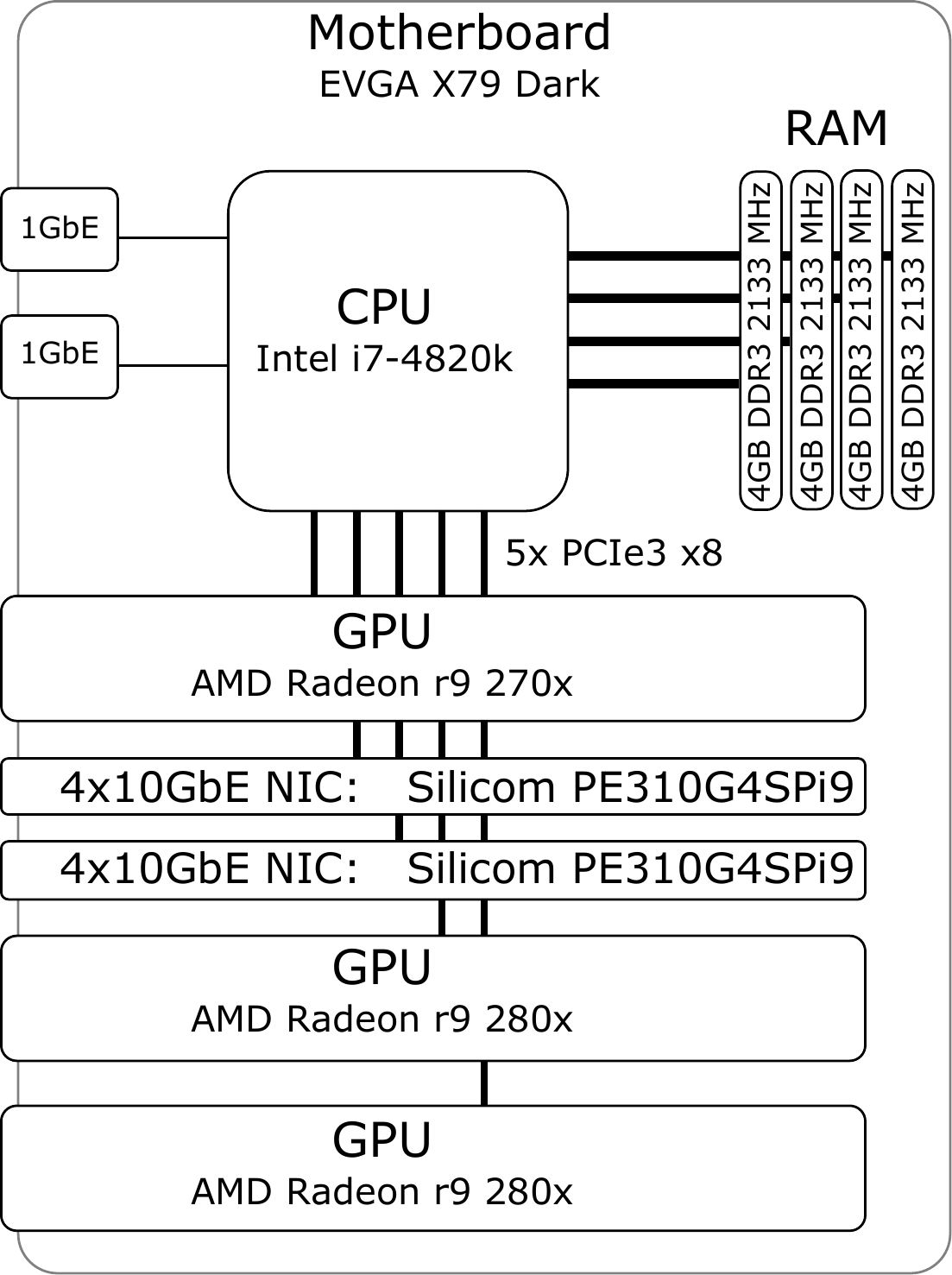}
	\caption{The internal components of the GPU nodes, including network interfaces, as described in \S\ref{sec:node}.}
	\label{fig_data}
\end{figure}

A schematic of the hardware layout in each GPU node is shown in Figure \ref{fig_data}.
The processing nodes are built on EVGA X79 Dark motherboards, Intel i7 4820k CPUs, and 4x4GB DDR3 RAM overclocked to 2133\,MHz.
They are powered by 1000\,W high-efficiency\footnote{80 Plus Platinum, corresponding to $\geq\!92\%$ efficiency at $50\%$ load}
power supplies and housed in 4U Chenbro RM41300-FS81 cases.
Each node is fed by eight 10\,GbE lines connected to a pair of Silicom PE310G4SPi9 quad-10GbE network interface cards.

Three AMD R9-series GPUs -- two 280X and one 270X -- receive and process a total of 25\,MHz of radio bandwidth from all the spatial inputs.
The 270X is not required for the pairwise correlation, but was added to ensure surplus computational power
was available to explore additional real-time processing and alternate correlation algorithms  (\S\ref{sec:gpu_algs}).

The nodes are diskless, both booting off and streaming data back to the {\tt gamelan} control node via onboard GbE, through a Cisco SRW2048-K9-NA switch.
A second on-board GbE connection allows for future expansion and data exchange between nodes,
not required in standard operation but available for planned algorithmic upgrades.

The average per-node power usage is $\sim$680\,W for the air-cooled and $\sim$630\,W for the liquid-cooled; the total system exceeds the 105\,TOPS requirement with a direct power consumption of $\approx$10\,kW.
The X-engine therefore achieves $\sim$11\,GOPS/W,
though we note that these are 4-bit arithmetic operations, not directly comparable to single-precision floating-point operations.

\subsection{Liquid Cooling Upgrade}
\label{sec:water}
In the summer of 2014, 10 of the GPU nodes were retrofitted to cool their CPU and GPUs via direct-to-chip liquid cooling systems.
This upgrade significantly eases the strain upon the existing A/C system, and is intended as a 
proving ground for the full CHIME correlator,
where traditional HVAC solutions become prohibitively expensive. 

Liquid cooling was implemented using aftermarket heatsinks attached to each GPU, and a Swiftech Apogee II combination pump/heatsink attached to the CPU.
Lab testing showed a reduction in GPU die temperatures to $10^\circ$C above the input water temperature,
with minimal dependence on the specific variety of heatsink.
Varieties from several vendors were ultimately deployed, with similar performance.
In operation, the watercooled GPUs maintained stable temperatures of 40-60$^\circ$C, compared to $\sim$85$^\circ$C for the air-cooled GPUs.
The cooler operating point and the removal of the low-efficiency GPU cooling fans significantly dropped the power draw per node with no impact on performance, and is expected to reduce the long-term failure rates.

\begin{figure}[!t]
	\centering
	\includegraphics[width=2.75in]{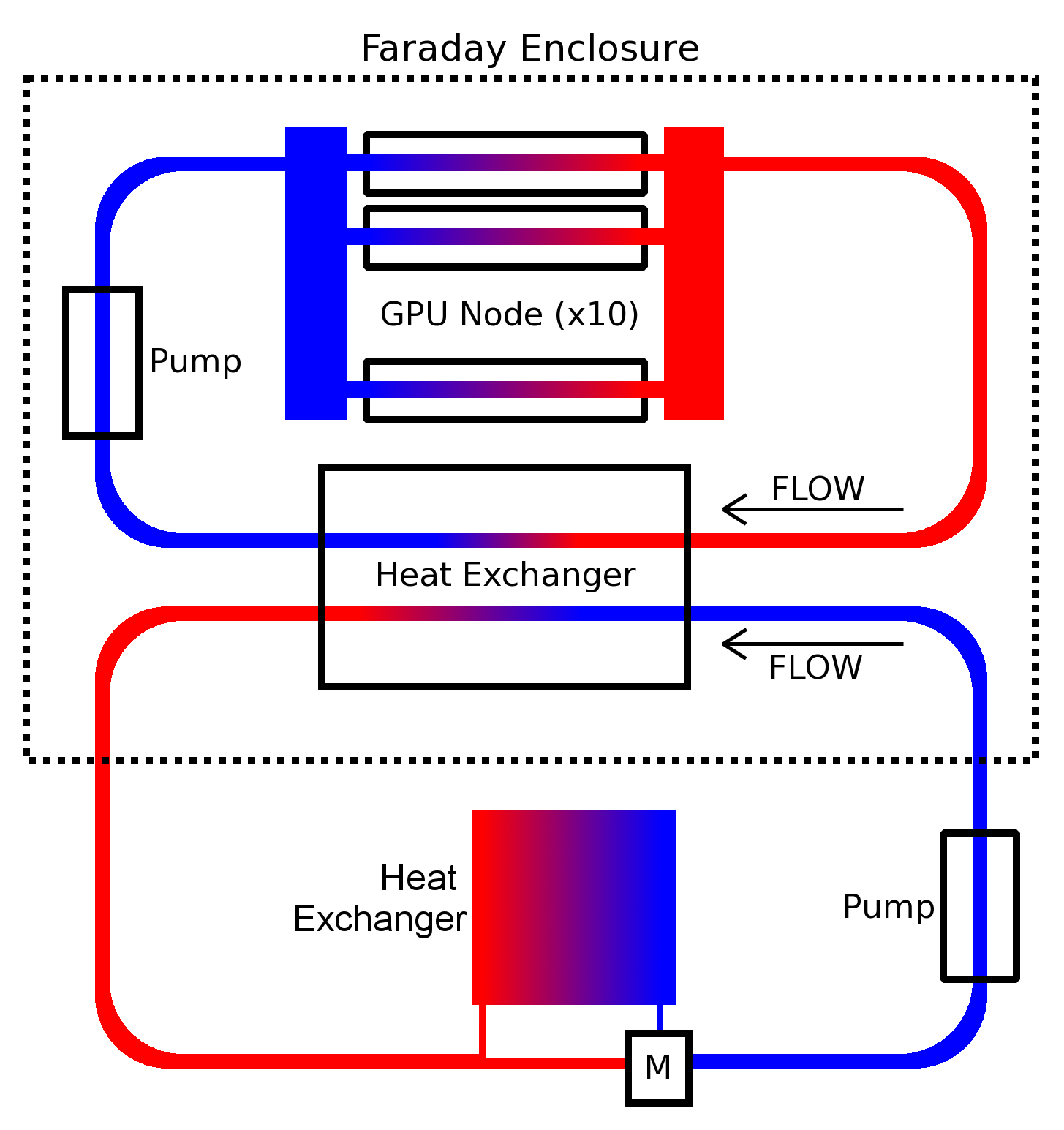}
	\caption{A diagram showing the liquid-cooling structure in the CHIME Pathfinder, with red and blue indicating hot and cold coolant, respectively. The heat exchanger uses a large fan to cool the liquid using ambient outside air. The object marked `M' is a temperature-controlled mixing valve, which regulates the temperature of the `cold' sections of the loop.}
	\label{fig:wcool}
\end{figure}

Figure \ref{fig:wcool} shows the overall structure of the watercooling system as currently deployed.
A sealed loop circulates a coolant consisting of 50\% water and 50\% ethylene glycol through the GPU nodes in parallel.
Each node has a small pump contained in the CPU heat sink and passive water blocks on each of the GPUs,
while a constant pressure differential is held across the nodes by a larger external pump.
The coolant runs through a heat exchanger which transfers heat from the node-cooling loop to an external liquid loop. 
That loop travels out of the RFI enclosure and is then cooled to the ambient external air temperature in a liquid-air heat exchanger.
A temperature-dependent mixing valve permits some fraction of the hot coolant to immediately recirculate,
to avoid cooling the system below the dew point inside the RFI enclosure during winter operation.

\section{Implementation - Software}
\label{sec:soft}

\subsection{Data Flow}

Channelized data, flags, and metadata from the F-engine arrive at the GPU nodes on eight 10\,GbE lines.
Each 10\,GbE link carries all spatial inputs for 8 of the 1024 total frequency channels.
With eight links per node, each node processes data covering $1/16$ of the full CHIME frequency band. 
The {\tt kotekan}\footnote{A style of playing fast interlocking parts in Balinese Gamelan music; see http://en.wikipedia.org/wiki/Kotekan}
software pipeline manages the data flow and processing within GPU nodes.   
Due to the high I/O demands ($\sim$820\,Gb/s in total),
the system must make maximally efficient use of the available bandwidth at each stage.
A packetized and loss-tolerant data handling system, similar to that in operation in the PAPER \cite{paper-32} correlator \cite{paper-corr}
ensures that momentary faults do not impede long-term data gathering.
Recnik et. al. \cite{recnik} discuss the data handling in detail; a brief description follows.

Data arrive as UDP packets and are buffered by the host CPU in system memory for inspection and staging prior to transfer into the GPUs for processing.
Packet loss, though rare, is tracked along with other flags from the F-engine from e.g. saturation of the ADCs or from FFT channelizing.
The count of missing or saturated data is used to renormalize the post-integration correlation matrices.

A series of OpenCL kernels are executed on the GPUs;
these pre-condition the data, compute and integrate correlation matrices,
and post-process the data if necessary (see \cite{klages} for more details).
Computed correlation matrices are assembled by the CPU and forwarded to {\tt gamelan}, which stitches the full 400\,MHz band back together
using data from all active nodes.
This reassembly is robust against individual node failures or outages; they simply result in loss of data from the inactive nodes.

Integrated correlation matrices are recorded onto an array of disks in {\tt gamelan}, and 
these data are asynchronously copied to a remote archive server hosting a much larger array of drives, and then copied off-site for scientific analyses.

\subsection{Monitoring and Control}

The X-engine software pipeline (composed of the {\tt kotekan} instances on each node,
along with the collection server software)
is launched and controlled through scripts run on {\tt gamelan}.
The nodes run CentOS Linux 6.5 and can be accessed by remote shell login,
while the PDUs allow remote power cycling to aid in recovery of crashed systems.

\begin{figure}
  \includegraphics[width=\columnwidth]{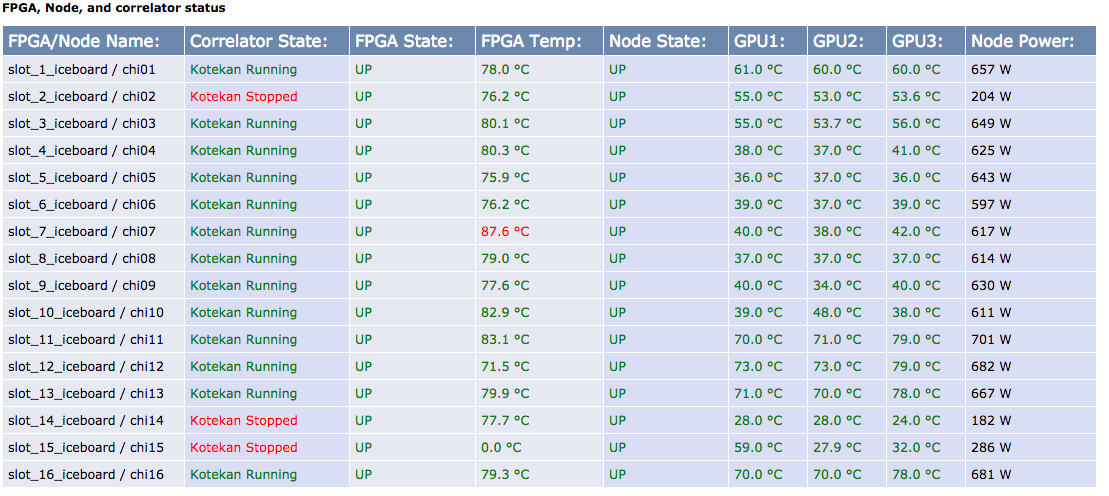}
  \caption{Screenshot showing the correlator status webpage.
	The monitoring system displays the status of each FPGA and GPU node, 
	and of the correlator software;
	per-GPU temperatures and per-node power consumption are also available.
	} 
  \label{fig:system_status}
\end{figure}

\begin{figure}
  \includegraphics[width=\columnwidth]{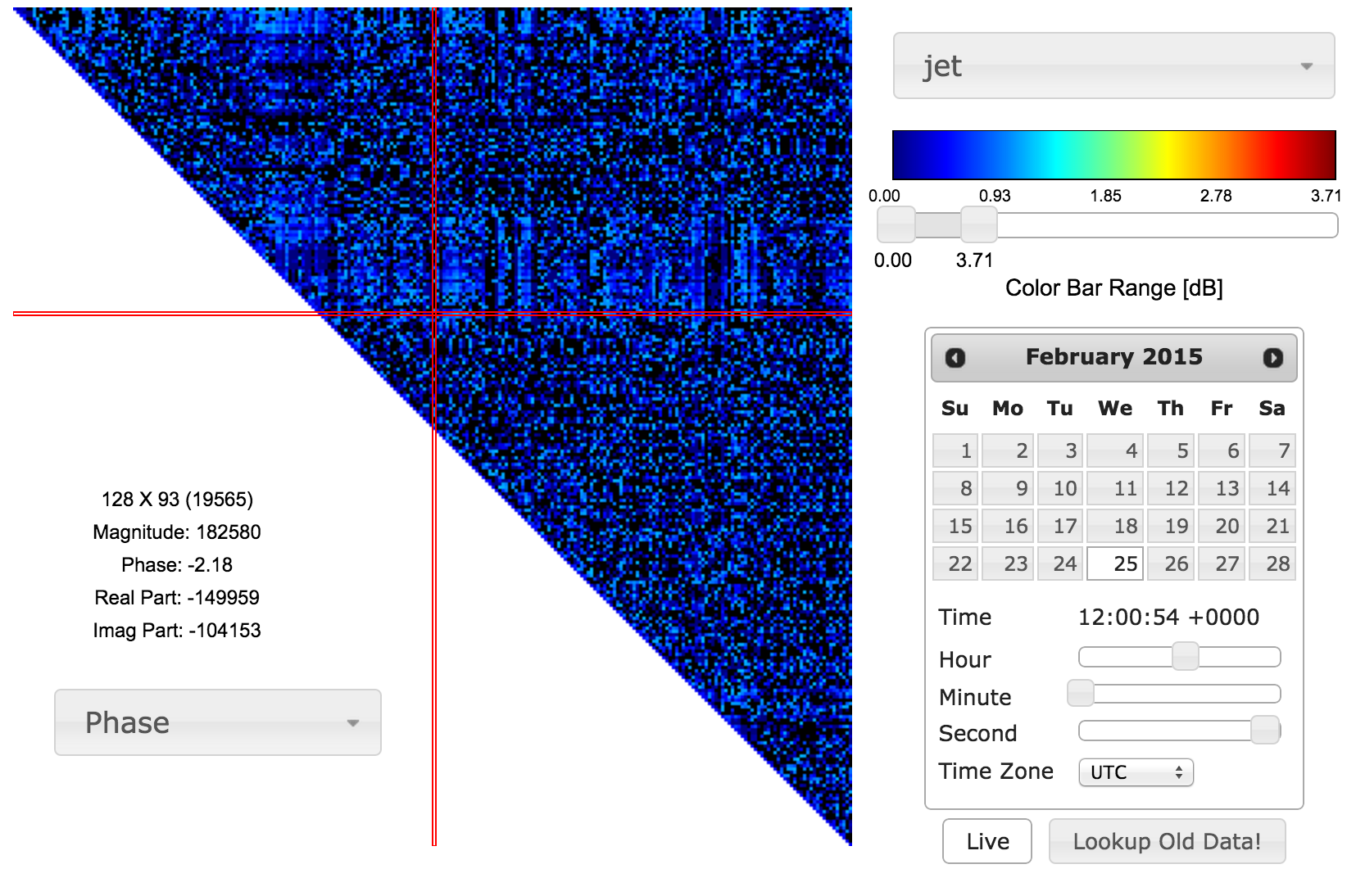}
  \caption{Screenshot showing the live view webpage. 
	The triangle is the full correlation matrix for a particular frequency, with colour indicating the complex value's phase; the website may be queried for any of the associated data.
	} 
  \label{fig:live_view}
\end{figure}
The status of each of the GPU nodes is tracked by the {\tt gamelan} control node, and made available via a web interface; see Figure~\ref{fig:system_status} for an example of the tracking display.

In addition, 
the last few hours of data are streamed over TCP to a second server, where it is available for live analysis and monitoring.
An example of the live-monitoring webpage is shown in Figure~\ref{fig:live_view}.

\subsection{GPU Data Processing Tasks}
\label{sec:gpu_algs}

The most computationally expensive operation performed on the GPU nodes is the mission-critical pairwise feed correlation.
In parallel with this, the Pathfinder correlator will explore alternate correlation methods which leverage the redundant layout of the CHIME baselines.
Supplemental tasks include beamforming, gating, time-shifting of inputs, and RFI excision. 
Brief descriptions of these tasks follow.

\subsubsection{Full Correlation}
\label{sec:crosscorr}
The primary responsibility of the X-engine is to calculate and integrate the correlation matrix of all the spatial inputs.
This involves accumulation of 32,896 pairwise products for each of 1024 frequency bands.
The default integration period is $2^{23}$ samples, corresponding to 21.47\,s, much faster than the $\gtrsim$2.5 minute beam-crossing time from sky rotation.
The computational requirements of the full X-engine system are dominated by this correlation operation, such that all other processes constitute an insignificant additional burden.
The current implementation achieves near-maximum-theoretical throughput; for details, see Klages et. al. \cite{klages}

\subsubsection{Alternate Correlation Techniques}
\label{sec:altcorr}
Interferometric arrays with highly redundant baselines can take advantage of correlation techniques that are more efficient than the na\"{\i}ve pairwise method. 
In the case of feeds which are evenly spaced, FFT-based transformations can be used to increase the efficiency of the correlation to $N\log N$, at the cost of strict calibration requirements. \cite{aag-red}\cite{TFFTT}\cite{tegtel}
These correlation strategies will be tested in parallel with the pairwise $N^{2}$ correlation; 
additionally, they may be used in hybrid form with some $N^2$ and some $N\log N$ stages.

\subsubsection{Discrete Beamforming}
\label{sec:beam}
The CHIME Pathfinder is a stationary telescope that cannot physically point at a specific source or location on the sky. 
When observing localized sources, it is desirable to form one or more beams, `pointing' the telescope digitally to an arbitrary location within the main beam.
This is accomplished in the GPUs by phase-shifting and summing the data from all antennas, so that signals originating in one region of the sky interfere constructively.
This signal is then written out at very high cadence, allowing examination of a localized source with very fine time resolution.

\subsubsection{Output Gating}
\label{sec:gating}
The CHIME Pathfinder will observe periodic sources such as astronomical pulsars and injected calibration signals. 
These sources generally vary faster than the default $\sim$21\,s integration period, but high-cadence gating may be used to observe sub-integration signal structure.
Gating consists of partitioning the output into a set of sub-buffers based on the time relative to the period of the source, so that independent `on' and `off' signals may be constructed.

\subsubsection{Time Shifting}
\label{sec:conditioning}
Signals from outlying telescope stations can be fed into the correlator.
Large spatial separations introduce decorrelation between inputs, which can be corrected for by time-shifting samples within the GPUs.
The current implementation permits the correction of any input by up to 168\,ms, and has been tested with the nearby John A. Galt 26m radio telescope at DRAO.

\subsubsection{RFI Cleaning}
\label{sec:cleaning}
Anthropogenic radio frequency interference (RFI) introduces a significant source of additional noise to the astronomical signal. 
These signals are generally narrow-band and intermittent, coming and going on timescales much shorter than the default 21\,s integration period,
but with relatively low duty cycles.
High-cadence identification and excision of RFI can be performed within the GPUs, and a variety of algorithms are under development 
including robust outlier and higher-moment statistical tests. \cite{sk-1}\cite{sk-2}

\section{X-engine Scalability}
\label{sec:scale}

The X-engine described here was designed for the CHIME Pathfinder, 
and must be scaled up significantly for the full CHIME instrument. 
Given a scaled F-engine providing channelized data,
the X-engine's design allows it to scale straightforwardly to a
broader band or larger-$N$ arrays.

Additional radio bandwidth is trivially added through additional nodes; 
increasing the number of inputs adds to the computational demand on each node,
and can be addressed through newer, more powerful GPUs.
At the time of writing, the computational power per node could be roughly tripled
by simply replacing the GPUs.

To support larger $N^2$ requirements, the bandwidth handled in each node can be reduced,
in exchange for proportionally more nodes.
The bandwidth fed to each GPU can similarly be reduced, and for very large $N$,
when even a single frequency band is beyond the capacity of a single processing node,
data can be time-multiplexed across multiple GPUs.

The expansion to full CHIME ($N=2048$) yields an $N^2$ computational requirement $\eta$
an order-of-magnitude greater than any system currently in existence.
Using {\em current} technology, a straightforward scaling of the current system
--- 256 nodes each containing 2 dual-chip R9 295X2 GPUs ---
could handle the entire pairwise correlation task, without additional software development.
This density of processors is easily achievable with the liquid cooling demonstrated,
and would occupy a modest physical footprint, at a very low hardware cost of $\sim$\$1M.
However, it is not expected that the full CHIME instrument will rely on a complete $N^2$ correlation, instead pursuing a fast alternate correlation technique as discussed in \S\ref{sec:altcorr}.

\section{Conclusion}
\label{sec:conc}

We have implemented a low-cost, high-efficiency GPU-based correlator X-engine for the CHIME Pathfinder.
Capable of correlating 32,896 baselines over 400\,MHz of radio bandwidth, 
it makes efficient use of consumer-grade parts and executes a highly optimized software stack.
Measured by the computational requirement of a na\"{\i}ve $N^2$ correlation -- the bandwidth-baseline product $\eta$ defined by Equation 1 --
the CHIME Pathfinder correlator is among the largest in the world.
Aspects of the system such as the cooling systems have been substantially modified, 
optimizing the X-engine's efficiency and ensuring economical scaling to the full-size CHIME instrument.

\section*{Acknowledgements}

We are very grateful for the warm reception and skillful help we have received from the staff of the Dominion Radio Astrophysical Observatory, which is operated by the National Research Council of Canada.

We acknowledge support from the Canada Foundation for Innovation, 
the Natural Sciences and Engineering Research Council of Canada, 
the B.C. Knowledge Development Fund,  
le Cofinancement gouvernement du Qu\'ebec-FCI, 
the Ontario Research Fund, 
the CIFAR Cosmology and Gravity program, 
the Canada Research Chairs program, and 
the National Research Council of Canada.
PK thanks IBM Canada for funding his research and work through the Southern Ontario Smart Computing Innovation Platform (SOSCIP).

We thank Xilinx University Programs for their generous support of the CHIME project,
and AMD for donation of test units.

\bibliographystyle{IEEEtran}
\bibliography{IEEEabrv,IEEEexample,report}
\end{document}